\documentclass[fleqn,twoside,twocolumn,nofootinbib,showkeys]{revtex4} % Specifies the document class %,unsortedaddress
\usepackage[sec,nocpr]{ujp} % \usepackage[cyr]{ujp} for cyrillic
\newcommand{\eq}[1]{\begin{equation} #1 \end{equation}}
\newcommand{\comment}[1]{}

\begin{document}
\title[Shear and Bulk Viscosities of the Hadron Gas]%колонтитул
{SHEAR AND BULK VISCOSITIES OF THE HADRON GAS WITHIN RELAXATION TIME APPROXIMATION AND ITS TEST}%
\author{O.~Moroz}%1 автор
\affiliation{Bogolyubov Institute for Theoretical Physics, Nat. Acad. of Sci. of Ukraine}%институт 1
\address{14b, Metrolohichna Str., Kyiv 03680, Ukraine}%адрес 1
\email{moroz@bitp.kiev.ua}%e-mail

\udk{539}  \razd{\secii}

\autorcol{O.\hspace*{0.7mm}Moroz}

\setcounter{page}{1}%

\begin{abstract}
We concentrate on calculation of the shear and bulk viscosities of the hadron gas. They define its dissipative dynamics and influence its experimentally measurable elliptic flow. Due to difficulty of this calculation the relaxation time approximation (RTA) was used in previous works. As those results have approached the realistic ones, there is a need to find out how accurate the RTA is. For this sake we calculate the viscosities in the RTA using cross sections extracted from the ultrarelativistic quantum molecular dynamics (UrQMD) model and compare them with the same ones calculated without the RTA. This allows us to find the estimates of errors due to the application of RTA in the calculations of the viscosities, which are valid also for other similar models. For instance, in the temperature region $100~MeV \lesssim T \lesssim 160~MeV$ at zero chemical potentials the shear viscosity becomes smaller up to $1.57$ times, or up to $1.45$ times if the averaged relaxation time is used. This has important consequences for interpretation of the previously made calculations of the viscosities and some other related calculations. Within the RTA, we also find estimation of the enhancement of the bulk viscosity of the hadron gas because of nonconservation of particle numbers. This is a little more extended version in compare to the published paper, where we were limited in paper's size and time.
\end{abstract}


\maketitle

\section{Introduction }

The bulk and shear viscosities are required for dissipative hydrodynamic description. This description finds applications to the strongly interacting matter created in heavy ion collisions. In particular, its elliptic flow can be measured, see review \cite{Kapusta:2008vb}.

In this paper we focus on calculation of the shear and bulk viscosity coefficients of the hadron gas at zero chemical potentials. Ref. \cite{Khvorostukhin:2010aj} provides these calculations, being close to the realistic ones, with unique advancement. However, they are done in the RTA.

The RTA for the Boltzmann equation (BE) is known for a long time, see, e. g., Ref. \cite{Reif}. The advantage of using it is that it provides simplification in calculations. However, in all its known realizations the errors from its application are not controlled. There are some tests and analysis of the RTA \cite{Plumari:2012ep, Wiranata:2012vv, Wiranata:2012br}, though it's not sufficient for estimation of the errors from RTA if the calculations for the hadron gas are required. Thus, there is a need to verify this approximation. When the error estimates for it are found, they can be used in other RTA-based calculations (directly or after modification, derivation or implementation of some ansatz), e. g., in Ref. \cite{Tawfik:2010mb} or hydro-kinetic calculations \cite{Sinyukov:2002if}. The heat conductivity and diffusion coefficients, being rather closer by their properties to the shear viscosity, could be expected to have approximately the same errors at the chemical potentials small compared to the temperature and with approximately the same densities.

It turns out that the bulk viscosity enhances much after introduction of inelastic (particle number changing) processes. Taking into account of them may have some difficulties, though in the RTA they can be taken into account relatively easily. One could also speak about approximate conservation of particle numbers, which has concrete mathematical realization for the bulk viscosity. Making comparison between the maximal and the minimal enhancements (or between the cases of minimal and maximal particle number conservations) within the same hadron gas model and within the same approximations, one could find the error estimates needed in Refs. \cite{Moroz:2013vd, Khvorostukhin:2010aj, Tawfik:2010mb}.

\section{The RTA and the Results }

The set of the BEs in the local rest frame in the RTA can be written as \cite{Reif}
%1
 \eq{
 \frac{df_k(t,\vec r,p_k^{0})}{dt}=-\frac{f_k(t,\vec r,p_k^{0})-f^{(0)}_k}{\tau_{\rm{rel},k}(t,\vec r,p_k^{0})},
 }
 %2
 \eq{
 f^{(0)}_k \equiv f^{(0)}_k(t,\vec r,p_k^{0}) = \exp(-p_k^0/T(t,\vec r)),
 }
 %3
 \eq{
 \tau_{\rm{rel},k}^{-1}(t,\vec r,p_k^{0})=\sum_{l}\int\frac{d^3p_{l}}{(2\pi)^3}f^{(0)}_{l} v_{kl} \sigma^{\rm{tot}}_{kl}(s),
 }
 \eq{
 v_{kl}\equiv \frac{\sqrt{ (s-m_k^2-m_l^2)^2-4m_k^2m_l^2}}{2 p_{k}^0p_{l}^0},
 }
where $f_k(t,\vec r,p_k^{0})$ and $f_k^{(0)}$ are nonequilibrium and local equilibrium distribution functions correspondingly, $\tau_{\rm{rel},k}(t,\vec r,p_k^{0}) \equiv \tau_{\rm{rel},k}(p_k^{0})$ is the relaxation time depending on the one-particle energy $p_k^0$ of the $k$-th species (cf. Ref. \cite{Chakraborty:2010fr}), $v_{kl}$ is the relativistic relative velocity (cf. Ref. \cite{Groot}), $\sigma^{\rm{tot}}_{kl}(s)$ is the total $2\leftrightarrow 2$ cross section (see also comments below), $s$ is the usual Mandelstam variable.

%Fig.1
\begin{figure}%
\vskip1mm
\includegraphics[width=\column]{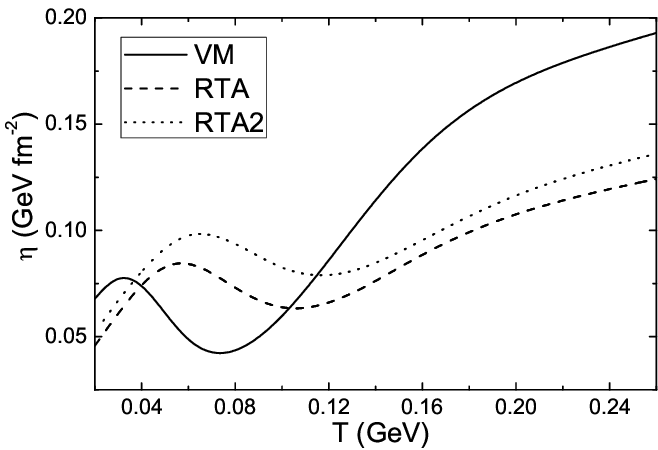}
\vskip-3mm\caption{ The shear viscosity versus temperature. The calculations are done using the variational method in the third order (solid line), using the RTA (dashed line) and using the RTA with the averaged relaxation time (dotted line).  \label{ShearViscFig} }

\end{figure}

%Fig.2
\begin{figure}%
\vskip3mm
\includegraphics[width=\column]{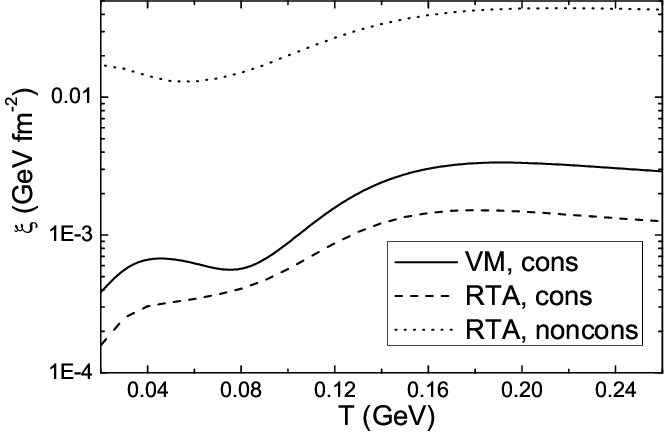}
\vskip-3mm\caption{ The bulk viscosity versus temperature. The calculations are done in the approximation of maximal conservation using the variational method in the fifth order (solid line) and using the RTA (dashed line). The calculations in the approximation of maximal nonconservation using the RTA are denoted by the dotted line. \label{BulkViscFig} }
\end{figure}

We also consider the momentum averaged relaxation time as in Ref. \cite{Khvorostukhin:2010aj}. In general, this approximation should not be better; however, we find it about as good as with the momentum dependent relaxation time in our calculations. In Ref. \cite{Khvorostukhin:2010aj} the transport cross sections as in Ref. \cite{Landau10} are used. The UrQMD hadronic cross sections \cite{Bass:1998ca, Bleicher:1999xi}, which we exploit\,\footnote{We use improved and extracted UrQMD cross sections as described in Ref. \cite{Moroz:2013vd}.}, have some extrapolation of the angular dependence from nucleons, but we do not just adopt it here. The expected deviations are of $4\%$ or less. With the improved transport cross sections as in Refs. \cite{Wiranata:2012vv, Plumari:2012ep} the isotropic total cross sections would get the $2/3$ extra factor (and the viscosities would get the factor $3/2$). This doesn't provide a better overall description of the viscosities, though it is better at high enough temperatures, see below.

We use the $\sigma^{\rm{tot}}_{kl}(s)$ as $2\leftrightarrow n$ total cross sections (TCSs), taking into account also quasielastic and other than $2\leftrightarrow 2$ processes, as in Ref. \cite{Moroz:2013vd}.
So that some cross sections add up exactly and some ones add up approximately into the total ones. This approximation is a good one, see Ref. \cite{Moroz:2013vd} for checkups. In addition to this, we also use the elastic plus quasielastic cross sections (EQCSs) \cite{Moroz:2013vd} and find approximately the same error estimates as with the TCSs. Other approximations which we apply (ideal gas equation of state, no medium effects, classical statistics) result in small corrections \cite{Moroz:2013vd, Khvorostukhin:2010aj} at least in the temperature range $100~{\rm MeV} \lesssim T \lesssim 160$~MeV at zero chemical potentials. If these corrections are not small, then in assumption of absent or weak correlations with the RTA corrections the latter ones can be yet \mbox{applicable.}

The relaxation time enters the shear $\eta$ and the bulk $\xi$ viscosities as (cf. Ref. \cite{Chakraborty:2010fr})
%4
 \eq{
 \eta=\frac1{15T}\sum_k\int \frac{d^3p_k}{(2\pi)^3} \frac{\tau_{\rm{rel},k}(p_k^{0})}{(p_k^0)^2}|\vec p|^4 f^{(0)}_{k},
 }
 %5
 \eq{
 \xi=T^3\sum_k\int \frac{d^3p_k}{(2\pi)^3} \frac{\tau_{\rm{rel},k}(p_k^{0})}{(p_k^0)^2}\hat Q_k^2f^{(0)}_{k},
 }
where $\hat Q_k$ is the dimensionless bulk viscosity source term which we take in the convenient form as in Ref. \cite{Moroz:2011vn}. There the approximation of maximal particle number conservation implies that the particle's charges are equal to the Kronecker's delta functions, $q_{ak}=\delta_{ak}$. The approximation of maximal nonconservation is equivalent to the case $q_{ak}=0$ at zero chemical potentials. There are also matching conditions, which can be satisfied modifying additionally the relaxation time \cite{Khvorostukhin:2012kw}. We do not investigate whether this modification of the RTA provides a better overall description. We are interested in testing the RTA as the one in Ref. \cite{Khvorostukhin:2010aj}.

Figs. \ref{ShearViscFig} and \ref{BulkViscFig} show the results of calculations at zero chemical potentials of the shear viscosity and the bulk viscosity correspondingly. In addition to the calculations in the RTA, there are also depicted calculations using the variational method with the application of the TCSs \cite{Moroz:2013vd}. From these results one can see the deviations from application of the RTA. To see how strong the dependence is of these deviations on the energy dependence of the TCSs, we also made the same calculations with the EQCSs. This almost does not change the deviations so that we do not show the results with the EQCSs. As long as the temperature dependence of the viscosities in the SHMC model \cite{Khvorostukhin:2010aj} and in the one of the present paper are similar to each other, the found error estimates from application of the RTA should be approximately the same for the SHMC model.

From Fig. \ref{ShearViscFig} one can see that in the important temperature range\,\footnote{At zero chemical potentials one has for the kinetic freeze-out temperature $T\approx 120$~MeV \cite{Heinz:2007in}, and for both the pseudo-critical temperature and the chemical freeze-out temperature one has $T\approx 160$~MeV \cite{Borsanyi:2010bp, Heinz:2007in}} $100~{\rm MeV} \lesssim T \lesssim 160$~MeV the shear viscosity becomes smaller up to $1.57$ times because of the application of the RTA. If the averaged relaxation time is used, these deviations are somewhat smaller and reach the factor 1.45, which is rather an accidental improvement. At smaller temperatures, in the vicinity of minimum of shear viscosity, these deviations are somewhat larger instead. Also the $3/2$ times larger shear viscosity in the RTA (see above) would give a better description at higher temperatures but a worse description in the vicinity of the minimum of shear viscosity. This minimum for the hadron gas is attributed to resonant peaks in the quasielastic cross sections of pions, dominating at those energies and temperatures. So this rapid change in the energy dependence of the cross sections does not permit the improved RTA description. At higher temperatures there are different cross section energy dependencies canceling each other approximately, which results in a better description with one constant cross \mbox{section \cite{Moroz:2013vd}.}\,\footnote{In this approximation the deviations from application of RTA are of the factor 1.6-1.7 in the whole considered temperature range.}

Fig. \ref{BulkViscFig} demonstrates the calculations of the bulk viscosity in the approximations of minimal and maximal particle number conservations (see comments above). The bulk viscosity calculations using the variational method are shown only in the approximation of the maximal conservation because only this one is considered in Ref. \cite{Moroz:2013vd}. As long as the calculations of the bulk viscosities using the RTA with the averaged and not averaged relaxation times differ by $13\%$ or less, we do not show the results with the averaged relaxation time. The bulk viscosity in the RTA turns out to be smaller at all the temperatures in 1.4--2.4 times\,\footnote{These numbers are replaced with 2.1-2.6 if the approximation of one constant cross-section is used.}.

Also one can see from Fig. \ref{BulkViscFig} that the enhancement of the bulk viscosity because of the maximal particle number nonconservation is large. At the chemical freeze-out temperature $T\approx 160$~MeV the ratio of the RTA-based bulk viscosity with the maximal nonconservation, $\xi_\text{noncons}$, to the one with the maximal conservation, $\xi_\text{cons}$, is equal to 27.27.
On the chemical freeze-out line, where the elastic plus the quasielastic rates, $\text{rate}_\text{elast}$, are equal to the total rates, $\text{rate}_\text{total}$, we do not know a priori what approximation dominates in the bulk viscosity, so that we need to divide this number by two to use it as an error estimate in the calculations with either of the approximations. Also neither of the approximations should be valid beyond its region delimited by the chemical freeze-out line. Assuming equal probabilities for both the boundary values, $\xi_\text{noncons}$ and $\xi_\text{cons}$, a little more accurate factor for the $\xi_\text{cons}$ would be (27.27+1)/2). Corresponding factor for the $\xi_\text{noncons}$ is easily reobtainable. At smaller temperatures (and the same chemical potentials) the same error estimate 13.64 or 14.14 (as approximate value or its upper bound) could be used because the inelastic processes become weaker there \cite{Bleicher:2002dm, Moroz:2013vd}. One could also connect the fading of the enhancement of the bulk viscosity to the collision rates and to multiply the number 27.27 by $1-{\rm rate}_{\,\rm elast}/{\rm rate}_{\,\rm total}$ (approximate formula for a limited temperature range too). More reasonable estimation of the factor would be $a(T) + b(T) x(T)$, where $x(T)=\text{rate}_\text{elast}/\text{rate}_\text{total}$ (can be taken in the form $\eta_\text{TCS2s}/\eta_\text{EQCS2s}$ from Ref. \cite{Moroz:2013vd}, being equal to 2.05 at the $T=160~MeV$), $a(T)=\xi_\text{noncons}/\xi_\text{cons}$, $b(T)=-\xi_\text{noncons}/\xi_\text{cons} + 1$. The temperatures $T\gtrsim 160$~MeV are less interesting because they are above the pseudo-critical temperature \cite{Borsanyi:2010bp}, however the mentioned estimates should be valid to some extent there too. We present these estimates with caveat because calculations based on the Chapman-Enskog and variational methods \cite{Moroz:2011vn} give notably larger estimate.

\section{ Conclusions }

In the direct comparison we have found that one can normally expect to have deviations in the viscosities from the application of the RTA up to 2--3 times with the energy dependence of cross sections as of the hadron gas at the considered temperatures $20~{\rm MeV} \leq T \leq 260$~MeV and zero chemical \mbox{potentials.}

At the temperatures $T\gtrsim 100$~MeV and zero chemical potentials the application of the RTA decreases the shear viscosity $\eta$. The factor of this deviation reaches 1.57 at $T=160$~MeV or 1.45 if the averaged relaxation time is used. As long as the temperature dependence of $\eta$ in the SHMC model \cite{Khvorostukhin:2010aj} and in the present paper are similar to each other, the found error estimates should be approximately the same there.
This confirms that multi-hadron production processes and some other ones, which are seemingly not taken into account in Ref. \cite{Khvorostukhin:2010aj}, are important to get the $\eta/s$ ($s$ is the entropy density) well consistent with the experimental data \cite{Moroz:2013vd}.

Using the TCSs, we have found that the ratio of the bulk viscosity with the maximal nonconservation to the one with the maximal conservation is equal to 27.27 at the chemical freeze-out temperature $T\approx 160$~MeV. This number should be divided by two on the chemical freeze-out line (approximate estimation), where neither conservation nor nonconservation is preferred a priori. At other temperatures some extrapolations could be used. For this goal further investigations are desirable. We present that estimate with caveat, because calculations based on the Chapman-Enskog and the variational methods \cite{Moroz:2011vn} give notably larger estimates.
If they are notably larger, this may be the case of the largest deviations from the application of the RTA.


\begin{thebibliography}{99}                                                                                                %
\bibitem{Kapusta:2008vb}
  J.~I.~Kapusta,
  %``Viscous Properties of Strongly Interacting Matter at High Temperature,''
  %"Relativistic Nuclear Collisions", Landolt-Bornstein New Series, Vol.\  I/23, ed.\  R.\ Stock {\bf } (Springer-Verlag, Berlin Heidelberg 2010).
  [arXiv:0809.3746 [nucl-th]].


\bibitem{Khvorostukhin:2010aj}
  A.S.~Khvorostukhin, V.D.~Toneev, and D.N.~Voskresensky,
  %``Viscosity Coefficients for Hadron and Quark-Gluon Phases,''
  Nucl. Phys. A {\bf 845}, 106 (2010)
  [arXiv:1003.3531 [nucl-th]].

\bibitem{Reif} F.~Reif, \emph{Fundamentals of Statistical and Thermal Physics}, (McGraw-Hill Book Company, New York, 1965), Chap. 13.



\bibitem{Plumari:2012ep}
  S.~Plumari, A.~Puglisi, F.~Scardina, and V.~Greco,
  %``Shear Viscosity of a strongly interacting system: Green-Kubo vs. Chapman-Enskog and Relaxation Time Approximation,''
  Phys. Rev. C {\bf 86}, 054902 (2012)
  [arXiv:1208.0481 [nucl-th]].

\bibitem{Wiranata:2012vv}
  A.~Wiranata, M.~Prakash, and P.~Chakraborty,
  %``Comparison of Viscosities from the Chapman-Enskog and Relaxation Time Methods,''
  Central Eur. J. Phys.  {\bf 10}, 1349 (2012)
  [arXiv:1201.3104 [nucl-th]].

\bibitem{Wiranata:2012br}
  A.~Wiranata and M.~Prakash,
  %``Shear Viscosities from the Chapman-Enskog and the Relaxation Time Approaches,''
  Phys. Rev. C {\bf 85}, 054908 (2012)
  [arXiv:1203.0281 [nucl-th]].

\bibitem{Tawfik:2010mb}
  A.~Tawfik and M.~Wahba,
  %``Bulk and Shear Viscosity in Hagedorn Fluid,''
  Annalen Phys.  {\bf 522}, 849 (2010)
  [arXiv:1005.3946 [hep-ph]].

\bibitem{Sinyukov:2002if}
  Y.M.~Sinyukov, S.V.~Akkelin, and Y.~Hama,
  %``On freezeout problem in hydro kinetic approach to A+A collisions,''
  Phys. Rev. Lett.  {\bf 89}, 052301 (2002)
  [nucl-th/0201015].

\bibitem{Moroz:2013vd}
  O.N.~Moroz,
  %``Analytical formulas, general properties and calculation of transport coefficients in the hadron gas: shear and bulk viscosities,''
  arXiv:1301.6670 [hep-ph].

\bibitem{Chakraborty:2010fr}
  P.~Chakraborty and J.I.~Kapusta,
  %``Quasi-Particle Theory of Shear and Bulk Viscosities of Hadronic Matter,''
  Phys. Rev. C {\bf 83}, 014906 (2011)
  [arXiv:1006.0257 [nucl-th]].

\bibitem{Groot}
S.R. de Groot, W.A. van Leeuwen and Ch.G. van Weert,
\emph{Relativistic Kinetic Theory} (North-Holland, Amsterdam, 1980),
Chap.~I, Sec.~2.

\bibitem{Landau10}
E.M. Lifschitz and L.P. Pitaevski, \emph{Physical kinetics}, 2. ed.
(Pergamon Press, Oxford, 1981), Sec. 11.

\bibitem{Bass:1998ca}
  S.A.~Bass, M.~Belkacem, M.~Bleicher, M.~Brandstetter, L.~Bravina, C.~Ernst, L.~Gerland, and M.~Hofmann {\it et al.},
  %``Microscopic models for ultrarelativistic heavy ion collisions,''
  Prog. Part. Nucl. Phys.  {\bf 41}, 225 (1998).

\bibitem{Bleicher:1999xi}
  M.~Bleicher, E.~Zabrodin, C.~Spieles, S.~A.~Bass, C.~Ernst, S.~Soff, L.~Bravina, and M.~Belkacem {\it et al.},
  %``Relativistic hadron hadron collisions in the ultrarelativistic quantum molecular dynamics model,''
  J. Phys. G {\bf 25}, 1859 (1999).

\bibitem{Moroz:2011vn}
  O.N. Moroz,
  %``Towards precise calculation of transport coefficients in the hadron gas. The shear and the bulk viscosities,''
  arXiv:1112.0277 [hep-ph].

\bibitem{Khvorostukhin:2012kw}
  A.S.~Khvorostukhin, V.D.~Toneev, and D.N.~Voskresensky,
  %``Relaxation time ansatz and bulk viscosity of hadron matter,''
  NPA 915, {\bf 159} (2013)
  [arXiv:1204.5855 [nucl-th]].

\bibitem{Heinz:2007in}
  U.W.~Heinz and G.~Kestin,
  %``Jozso's Legacy: Chemical and Kinetic Freeze-out in Heavy-Ion Collisions,''
  Eur. Phys. J. ST {\bf 155}, 75 (2008).

\bibitem{Borsanyi:2010bp}
  S.~Borsanyi {\it et al.}  [Wuppertal-Budapest Collaboration],
  %``Is there still any T_c mystery in lattice QCD? Results with physical masses in the continuum limit III,''
  JHEP {\bf 1009}, 073 (2010).

\bibitem{Bleicher:2002dm}
  M.~Bleicher and J.~Aichelin,
  %``Strange resonance production: Probing chemical and thermal freezeout in relativistic heavy ion collisions,''
  Phys. Lett. B {\bf 530}, 81 (2002).


\begin{flushright}
{\footnotesize Received 25.10.13}
\end{flushright}
\end{thebibliography}
\end{document}